\documentclass[pra,twocolumn,showpacs,aps,floatfix,amsmath,amssymb,amsfonts]{revtex4-1}
\usepackage{graphicx}
\usepackage{amsmath,amssymb,revsymb}
\usepackage{color}
\usepackage[colorlinks=true,citecolor=blue]{hyperref}

\begin{document}

\title{Contrasting the wide Feshbach resonances in $^6$Li and $^7$Li}

\author{Paul S. Julienne}
\affiliation{Joint Quantum Institute (JQI), University of Maryland and NIST, College
Park, Maryland, 20742, USA}
\author{Jeremy M. Hutson}
\affiliation{Joint Quantum Centre (JQC) Durham/Newcastle, Department of
Chemistry, Durham University, South Road, Durham, DH1~3LE, United
Kingdom}

\begin{abstract}
We compare and contrast the wide Feshbach resonances and the corresponding
weakly bound states in the lowest scattering channels of ultracold $^6$Li and
$^7$Li. We use high-precision measurements of binding energies and scattering
properties to determine interaction potentials that incorporate
non-Born-Oppenheimer terms to account for the failure of mass scaling between
$^6$Li and $^7$Li. Correction terms are needed for both the singlet and the
triplet potential curves. The universal formula relating binding energy to
scattering length is not accurate for either system. The $^6$Li resonance is
open-channel-dominated and the van der Waals formula of Gao [J. Phys.\ B 37,
4273 (2004)] gives accurate results for the binding energies across much of the
resonance width. The $^7$Li resonance, by contrast, is weakly
closed-channel-dominated and a coupled-channels treatment of the binding
energies is required. Plotting the binding energies in universal van der Waals
form helps illustrate subtle differences between the experimental results and
different theoretical forms near the resonance pole.
\end{abstract}

\date{\today}


\maketitle

\section{Introduction}

The magnetically tunable threshold scattering resonances of two atoms provide a
powerful tool for investigating many-body and few-body phenomena in ultracold
quantum gases~\cite{Chin2010}.  Here we compare and contrast the wide
resonances of the species $^6$Li and $^7$Li using accurate quantum scattering
calculations interpreted within the framework of universal van der Waals
quantum defect theory.  These Li resonances, which are quite different in
character despite a superficial similarity in magnetic-field
width~\cite{Chin2010}, have been used in numerous experimental studies
involving the $^6$Li fermion~\cite{Cubizolles2003, Jochim2003b, Strecker2003,
Bourdel2004, Zwierlein2004, Bartenstein2005, Schunck2005, Partridge2005,
Zwierlein2005, Zwierlein2006, Partridge2006, Zurn2013} or $^7$Li
boson~\cite{Abraham1997, Bradley1997, Strecker2002, Pollack2009, Gross2009,
Gross2010, Gross2011, Rem2013, Dyke2013}. They serve as a prototype of the
variations encountered among the many different resonances and species of
interest for ultracold physics.  Using the universal properties of the van der
Waals potential gives a powerful way to characterize the variation in resonance
properties in terms of dimensionless variables.

The fundamental quantity for studies of the interactions of ultracold atoms at
very small collision energy $E$ is the $s$-wave scattering length, which can be
tuned approximately according to the resonant formula~\cite{Moerdijk1995,
Chin2010}
\begin{equation}
 a(B) = a_\mathrm{bg} \left ( 1 - \frac{\Delta}{B-B_0} \right ) \,, \label{aB}
\end{equation}
where $a_\mathrm{bg}$ is a near-constant ``background'' scattering length far
from the resonance pole at magnetic field $B=B_0$ and $\Delta$ is the resonance
width. The resonance is due to the variation with magnetic field of the energy
of a ``closed-channel'' bound state with a magnetic moment $\mu_\mathrm{mol}$
that is different from the combined magnetic moment $\mu_\mathrm{atoms}$ of the
two separated atoms that define the open entrance channel. The mixing of the
``bare'' or uncoupled open and closed channels results in a coupled-channels
bound state with an energy $E_\mathrm{b}$ that is universally related to the
scattering length when the latter is sufficiently large~\cite{Chin2010},
\begin{equation}
  E_\mathrm{b}(B) \approx -\frac{\hbar^2}{2\mu a(B)^2} \,, \label{UE}
\end{equation}
where $\mu$ is the reduced mass of the pair of atoms.  While this equation is
widely used, it is actually not quantitatively very accurate until the
scattering length become extraordinarily large, and departures from it show up
readily in experimental measurements of binding energies.  In fact, the binding
energy of a Feshbach molecule comprised of two $^6$Li atoms in different spin
states has been measured so accurately~\cite{Zurn2013} that the actual binding
energy of the resonant state deviates from the universal value in
Eq.~(\ref{UE}) by 200 times the measurement uncertainty, even when the
scattering length is on the order of 2000 $a_0$, where $a_0$ is the Bohr
radius.  We will use accurate coupled-channels calculations to investigate the
relationship between $E_\mathrm{b}(B)$ and $a(B)$ for resonances for both
$^6$Li atoms and $^7$Li atoms, and compare these to the predictions of simple
single-channel formulas that correct Eq.~(\ref{UE}) in the case of a van der
Waals potential~\cite{Gribakin1993,Gao2004b,Chin2010}.  The two species show
quite different departures from the universal formula because of their very
different resonance character.

The long-range potential between two S-state neutral atoms has the van der
Waals form $-C_6/R^6$, where $R$ is the interatomic distance. The strength of
the van der Waals potential sets a characteristic length and energy associated
with low-energy collisions~\cite{Gribakin1993,Chin2010},
\begin{equation}
  \bar{a} = \frac{2\pi}{\Gamma(\frac14)^2}
            \left (  \frac{2\mu C_6}{\hbar^2} \right )^\frac14;\qquad
  \bar{E} = \frac{\hbar^2}{2\mu \bar{a}^2}. \label{baraE}
\end{equation}
Using $C_6 =$ 1393.39 E$_\mathrm{h}a_0^6$~\cite{Yan1996}, where $E_{\rm h}$ is
the Hartree energy, and the respective values of $\bar{a}$ and $\bar{E}/h$ are
29.884 $a_0$ and 671.93 MHz for two $^6$Li atoms and 31.056 $a_0$ and 533.41
MHz for two $^7$Li atoms.

Ultracold $s$-wave collisions occur in the domain of collision energy $E \ll
\bar{E}$ or, correspondingly, of de Broglie wavelength $2\pi/k \gg \bar{a}$,
where $\hbar k$ is the relative collision momentum. Furthermore, the ``pole
strength'' of the resonant pole in $a(B)$ in Eq.\ (\ref{aB}) is characterized
by a dimensionless parameter $s_\mathrm{res}=(a_\mathrm{bg}/\bar{a})(\delta \mu
\Delta/\bar{E})$, where $\delta \mu=\mu_\mathrm{atoms}-\mu_\mathrm{mol}$.  Chin
{\it et al.}~\cite{Chin2010} distinguish two distinct types of resonance. Those
with $s_\mathrm{res} \gg 1$ are ``open-channel-dominated'' and have a bound
state that takes on the character of the open entrance channel over a tuning
range spanning much of the width of the resonance.  By contrast,
``closed-channel-dominated'' resonances with $s_\mathrm{res} \ll 1$ have bound
states that take on the character of the closed channel except when $B$ is
tuned very close to the resonance pole.  The $^6$Li and $^7$Li resonances that
we study have respective $s_\mathrm{res}$ parameters of 59~\cite{Chin2010} and
0.49~\cite{Jachymski2013}, and thus show very different relationships between
$E_\mathrm{b}(B)$ and $a(B)$, in spite of the fact that they have widths
$\Delta$ of similar magnitude.

For heavy diatomic molecules, the differences between the binding energies for
different isotopes can be well accounted for by retaining the same potential
curves and simply changing the masses used in the calculation ~\cite{Seto:2000,
Docenko:2007, Falke:2008, Kitagawa2008, Strauss2010, Brue:AlkYb:2013,
Borkowski2013}. However, it is known that mass corrections due to the breakdown
of the Born-Oppenheimer approximation are important for light species like
H$_2$~\cite{Kolos1964, Kolos1965}, and also have significant effects in LiK
\cite{Tiemann:2009} and LiRb \cite{Ivanova:2011}. For the case of Li$_2$, Le
Roy and coworkers have analyzed extensive electronic spectra and have shown
that mass corrections are essential for both the singlet~\cite{Leroy2009} and
triplet~\cite{Dattani2011} states. In order to reproduce the resonance
positions, we also find that we have to use slightly different singlet and
triplet potentials for the two isotopes. The mass corrections correspond to an
isotopic shift of about 4~G in the position of the $^7$Li resonance from its
mass-scaled position.

This paper will describe the basic molecular physics of the near-threshold
states of the Li$_2$ molecule and our fitting of the potentials to the combined
experimental results for $^6$Li$_2$ and $^7$Li$_2$. Our potentials reproduce
the measured threshold two-body results for both isotopes. We will then compare
the binding energies calculated from our coupled-channels models, using a
universal van der Waals form to demonstrate the near-threshold relationships
between scattering length and binding energy and to test the approximate
formulas that have been developed to treat this relationship.  Both isotopes
exhibit clear deviation from the universal predictions of Eq.~(\ref{UE}), even
in regions of magnetic tuning where the scattering length is large compared to
$\bar{a}$. However, the behavior depends upon the value of the $s_\mathrm{res}$
parameter. We will show that the formula of Gao~\cite{Gao2004b} for the bound
states in a single van der Waals potential provides a good approximation for
the strong resonance in $^6$Li$_2$ but fails to represent the binding energies
for the much weaker resonance in $^7$Li$_2$.

\section{Overview of $^6$Li and $^7$Li}

\begin{figure}[tb]
\vspace{-0.8cm}
\includegraphics[width=\columnwidth,clip]{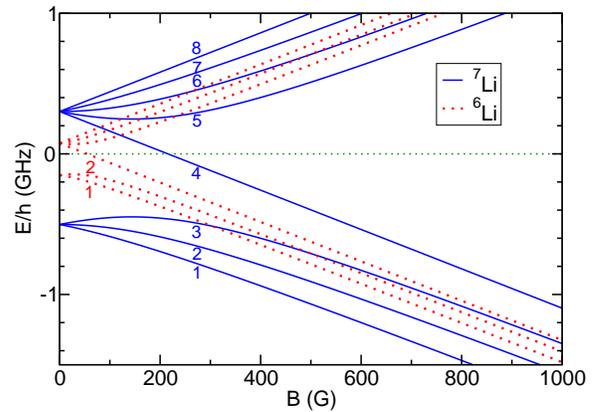}
\caption{(Color online) Energy levels of the $^2$S$_{1/2}$ ground state of the
$^6$Li and $^7$Li atoms, showing the different hyperfine structure of the two
isotopes, which at zero field have upper and lower total angular momentum quantum
numbers $f$=$\frac12$ and $\frac32$ for $^6$Li and $f=$  1 and 2 for $^7$Li.
The energy zero is the energetic center of gravity of the hyperfine multiplet.
The numbers indicate the state labels for all the states of $^7$Li and the
lowest two states of $^6$Li.  The lowest and highest-energy states at finite
$B$ have $m_f$=$+\frac12$ and $+\frac32$ for $^6$Li and $m_f$=+1 and +2 for
$^7$Li.} \label{fig:EvsBatom}
\end{figure}

Figure~\ref{fig:EvsBatom} shows the hyperfine/Zeeman atomic structure of the
$^2$S$_{1/2}$ ground state of the $^6$Li and $^7$Li atoms. The electron spin
$s=\frac12$ couples to the nuclear spin $i$ to give a resultant atomic spin $f$
with projection $m_f$.  When $B$ is non-zero, the states of different $f$ mix,
but the projection $m_f$ remains a good quantum number.  We designate the
atomic states in order of increasing energy by the labels 1, 2, $\ldots$ for
each species. A collision of two atoms is characterized by their relative
angular momentum given by the partial-wave quantum number $L$.  We are
interested in the lowest-energy $L=0$ ($s$-wave) spin channels of $^6$Li$_2$
and $^7$Li$_2$, for which the Feshbach resonances have been identified and
characterized. These are the (1,2) channel of $^6$Li$_2$~\cite{Schunck2005,
Bartenstein2005, Zurn2013} with $M_F=m_{f,{\rm a}}+m_{f,{\rm b}} = 0$ and the
(1,1)~\cite{Pollack2009, Gross2010, Gross2011, Rem2013, Dyke2013} and
(2,2)~\cite{Gross2009, Gross2011} channels of $^7$Li$_2$ with $M_F=$ 2 and 0
respectively.  Here we label the channels by the states $(i,j)$ of the two
separated atoms.

\begin{figure}[tb]
\vspace{-0.8cm}
\includegraphics[width=\columnwidth,clip]{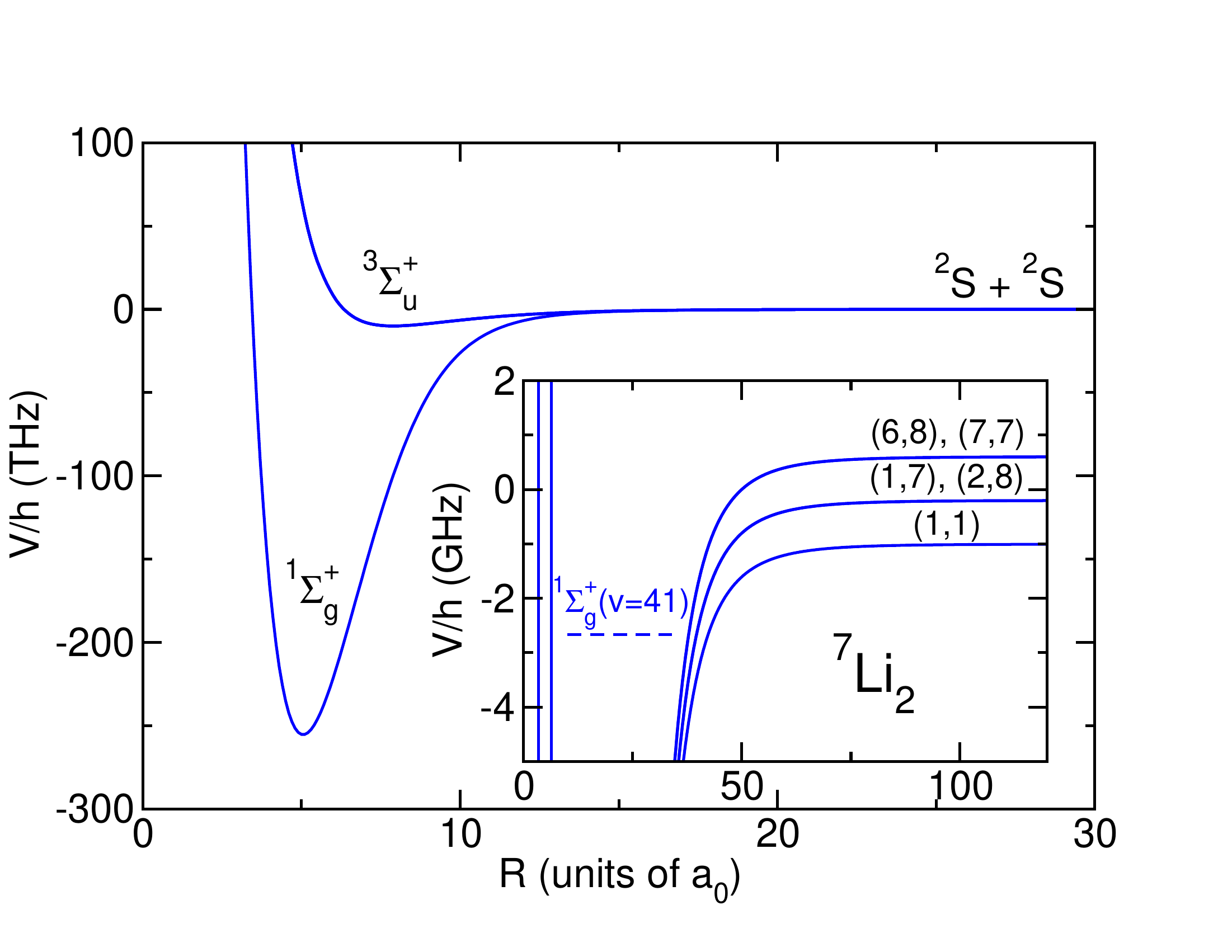}
\caption{(Color online) Born-Oppenheimer potential energy curves of the Li$_2$
molecule correlating with two ground-state atoms.  The inset shows the
adiabatic potential curves of the 5-channel potential matrix at $B=0$ for the
states with partial wave $L=0$ ($s$-wave) and total spin projection
$M_\mathrm{tot}=2$, using the notation of Fig.~\ref{fig:EvsBatom} to label the
channels. The inset also shows the location of the last zero-field $s$-wave
level of the $^7$Li$_2$ molecule.} \label{fig:Li2VR}
\end{figure}

Figure~\ref{fig:Li2VR} shows the adiabatic potential energy curves for the
lowest singlet and triplet states of the Li$_2$ molecule, $^1\Sigma_g^+$ and
$^3\Sigma_u^+$. The inset shows the asymptotic hyperfine structure for the five
spin channels of $^7$Li$_2$ that have projection $M_F=2$, which are (1,1),
(1,7), (2,8), (6,8) and (7,7). For a light species like Li, where the spacing
between vibrational levels is much larger than the atomic hyperfine splitting,
the molecular bound states, even near threshold, are primarily of either
$^1\Sigma_g^+$ or $^3\Sigma_u^+$ character. The inset of Fig.~\ref{fig:Li2VR}
indicates the energy $-2.668$ GHz $\times\ h$ of the last zero-field $s$-wave
bound state of the $^7$Li$_2$ molecule; this is the $v=41$ vibrational level of
the $^1\Sigma_g^+$ potential, with total nuclear spin $I=2$ and projection
$M_I=2$. The next $s$-wave bound states down from threshold are the hyperfine
components of a $^3\Sigma_u^+$ level near $-12$ GHz.  A similar figure for
$^6$Li in Ref.~\cite{Chin2010} shows the long-range hyperfine channels for the
$M_F=0$ states of $^6$Li$_2$; in this case the highest zero-field level is the
$v=38$ vibrational level of the $^1\Sigma_g^+$ state, which has two nuclear
spin components $I=0$, $M_I=0$ and $I=2$, $M_I=0$ that lie respectively at
$-1.625$ GHz and $-1.612$ GHz. The next $^6$Li$_2$ levels down are components
of the $^3\Sigma_u^+$ state near $-24$ GHz.

\begin{figure}[tb]
\vspace{-0.8cm}
\includegraphics[width=\columnwidth,clip]{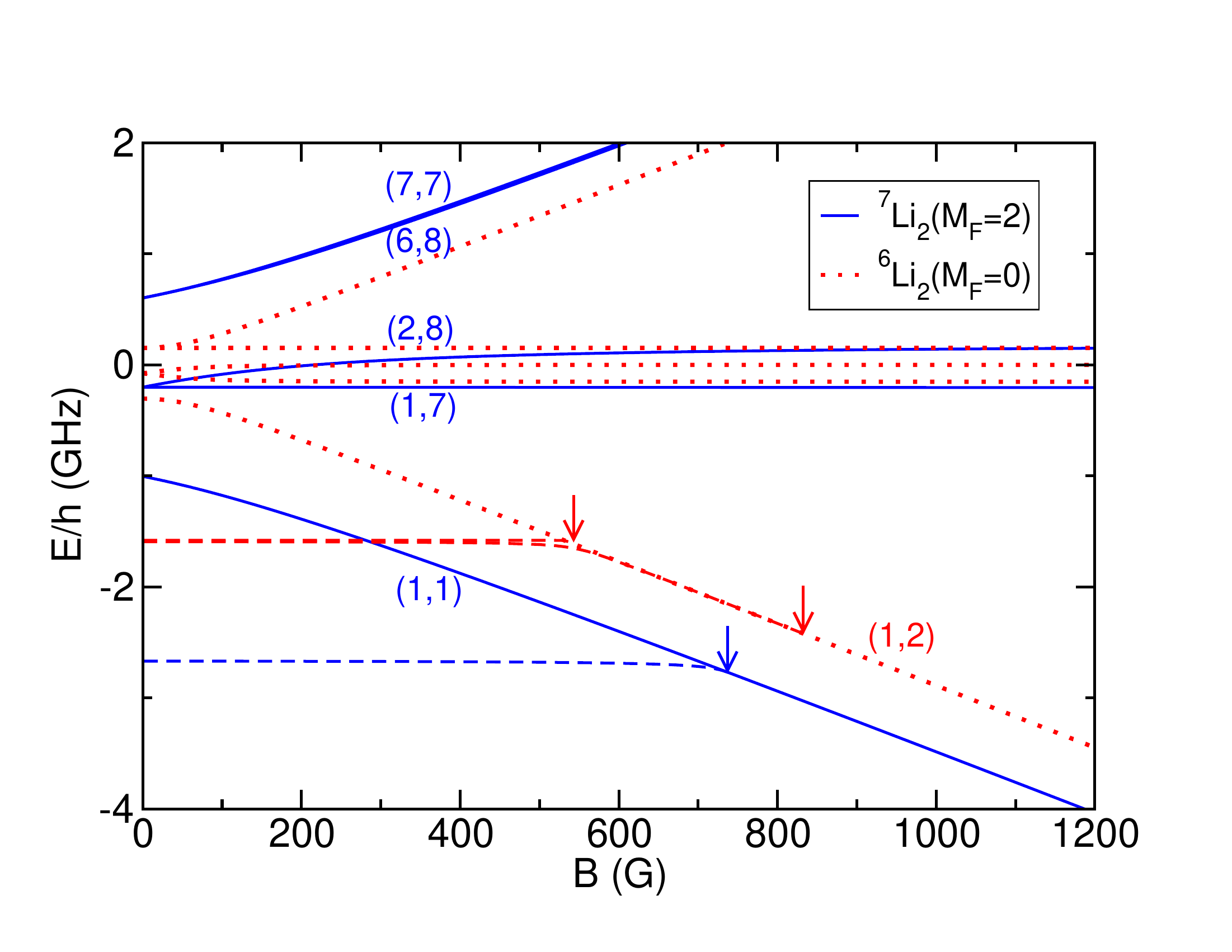}
\caption{(Color online) Energies of the separated-atom channels for $s$-wave
scattering with total spin projection $M_F=2$ for
$^7$Li$_2$ (solid lines) and $M_F=0$ for $^6$Li$_2$ (dotted lines). The dashed
lines show the energy of the last molecular bound state in each case.   The
arrows show the pole positions $B_0$ of the Feshbach resonances in the
lowest-energy $s$-wave channel of each species. } \label{fig:EvsBmol}
\end{figure}
Figure~\ref{fig:EvsBmol} shows the separated-atom energies for the respective
$M_F=0$ and $M_F=2$ states of $^6$Li$_2$ and $^7$Li$_2$. The upper dashed line
for $^6$Li$_2$ shows the $v=38$, $I=2$, $M_I=0$ $^1\Sigma_g^+$ level that is
very weakly coupled to the entrance-channel continuum.  It crosses threshold
near 543.26 G to make a very narrow closed-channel-dominated
resonance~\cite{Strecker2003, Schunck2005, Chin2010, Hazlett2012}.  The lower
dashed line shows the bound state that makes the very broad
open-channel-dominated resonance near 832~G~\cite{Bartenstein2005, Zurn2013}.
For fields below approximately 540~G, this bound state has the character of the
$v=38$, $I=0$, $M_I=0$ $^1\Sigma_g^+$ level, but it switches at higher $B$ to
become the last ($v=10$) level of the $^3\Sigma_u^+$ state, with the spin
character of the $(1,2)$ channel.  Above about 600~G, it is a halo state of
open-channel character and produces a scattering length that is large compared
to the van der Waals length for a field range spanning nearly 200~G below
resonance, or approximately 70\% of the resonance width $\Delta \approx$ 300
G~\cite{Chin2010}.  While the $v=41$, $I=2$, $M_I=2$ $^1\Sigma_g^+$ level of
$^7$Li shown in Fig.~\ref{fig:EvsBmol} also produces a resonance with a large
width $\Delta \approx$ 170 G~\cite{Gross2011, Dyke2013}, in this case the bound
state takes on the character of an open-channel halo molecule over only
approximately 1\% of the resonance width, very close to the resonance pole. We
will demonstrate that the very different character of the $^6$Li and $^7$Li
resonances shows up clearly in precise measurements of binding energies.

\begin{figure}[tb]
\includegraphics[width=\columnwidth,clip]{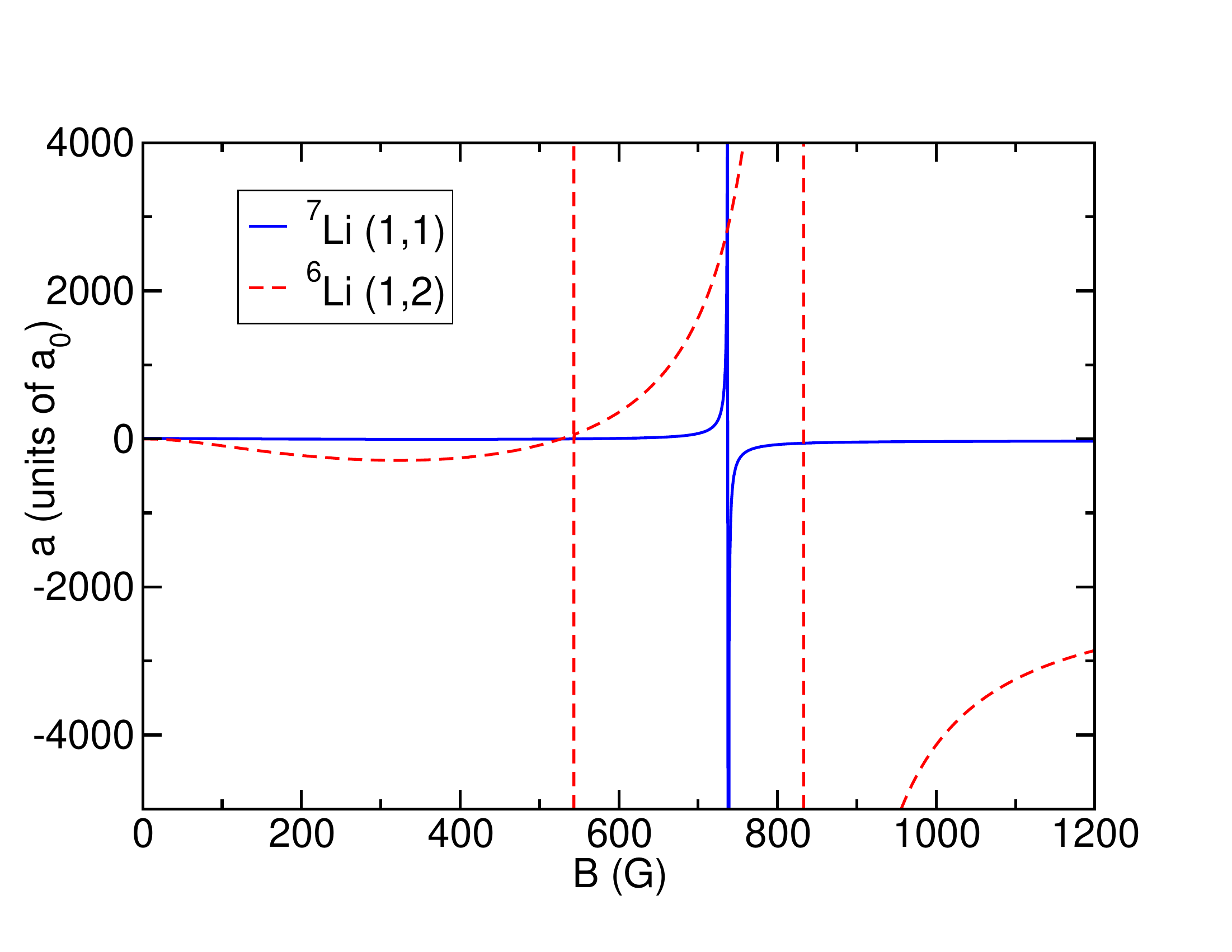}
\caption{(Color online) Scattering length $a$ as a function of magnetic field
$B$ for the (1,2) channel of $^6$Li and the (1,1) channel of $^7$Li, showing
the difference in pole strength for the two systems despite a superficial
similarity in magnetic-field width $\Delta$. The feature near 543~G in $^6$Li
is an additional narrow resonance.} \label{fig:scat-len}
\end{figure}

The scattering lengths for $^6$Li and $^7$Li are shown as a function of
magnetic field in Fig.\ \ref{fig:scat-len}. Despite the similarity in
magnetic-field widths $\Delta$, it may be seen that they have visually quite
different pole strengths corresponding to their different $s_{\rm res}$ values.
The magnetic-field width $\Delta$ for $^7$Li is in a sense anomalously large
because the resonance has a very small background scattering length.

\section{Theoretical model}
\label{Theory}

The Hamiltonian for the interaction of two alkali-metal atoms in their ground
$^2S$ states may be written
\begin{equation}
\frac{\hbar^2}{2\mu} \left[-R^{-1} \frac{d^2}{dR^2} R + \frac{\hat
L^2}{R^2} \right] + \hat h_1 + \hat h_2 + \hat V(R),
\label{eq:SE}
\end{equation}
where $\hat L^2$ is the operator for the end-over-end angular momentum of the
two atoms about one another, $\hat h_1$ and $\hat h_2$ are the monomer
Hamiltonians, including hyperfine couplings and Zeeman terms, and $\hat V(R)$
is the interaction operator.

In the present work we solve the bound-state and scattering problems by
coupled-channels calculations using the MOLSCAT \cite{molscat:v14} and BOUND
\cite{Hutson:bound:1993} packages, as modified to handle magnetic fields
\cite{Gonzalez-Martinez:2007}. Both scattering and bound-state calculations use
propagation methods and do not rely on basis sets in the interatomic distance
coordinate $R$. The methodology is exactly the same as described for Cs in
Section IV of Ref.\ \cite{Berninger:Cs2:2013}, so will not be repeated here.
The basis sets included all functions for $L=0$ and $L=2$ with the required
$M_{\rm tot}=M_F+M_L$. The energy-dependent $s$-wave scattering length $a(k)$
is obtained from the diagonal S-matrix element in the incoming channel,
\begin{equation}
\label{theory:eq5}
a(k) = \frac{1}{ik} \left(\frac{1-S_{00}}{1+S_{00}}\right),
\end{equation}
where $k^2=2\mu E/\hbar^2$ and $E$ is the kinetic energy
\cite{Hutson:res:2007}.

The interaction operator $\hat V(R)$ may be written
\begin{equation}
{\hat V}(R) = \hat V^{\rm c}(R) + \hat V^{\rm d}(R).
\label{eq:V-hat}
\end{equation}
Here $\hat V^{\rm c}(R)=V_0(R)\hat{\cal{P}}^{(0)} + V_1(R)\hat{\cal{P}}^{(1)}$
is an isotropic potential operator that depends on the electronic potential
energy curves $V_0(R)$ and $V_1(R)$ for the lowest singlet and triplet states
of Li$_2$, as shown in Figure~\ref{fig:Li2VR}. The singlet and triplet
projectors $\hat{ \cal{P}}^{(0)}$ and $\hat{ \cal{P}}^{(1)}$ project onto
subspaces with total electron spin quantum numbers 0 and 1 respectively. The
term $\hat V^{\rm d}(R)$ accounts for the dipolar interaction between the
magnetic moments of the two atoms, and for Li is represented simply as
\begin{equation}
\label{eq:Vd} \hat V^{\rm d}(R) = \frac{E_{\rm h} \alpha^2}{(R/a_0)^3} \left [
\hat s_1\cdot \hat s_2 -3 (\hat s_1 \cdot \vec e_R)(\hat s_2 \cdot \vec e_R)
\right ] \,,
\end{equation}
where $\vec e_R$ is a unit vector along the internuclear axis and
$\alpha\approx 1/137$ is the fine-structure constant.

At long range, the electronic potentials are
\begin{eqnarray}
V_S^{\rm LR}(R) = &-& C_6/R^6 - C_8/R^8 - C_{10}/R^{10}
\nonumber\\
&\pm& V_{\rm ex}(R),
\label{eq:vlr}
\end{eqnarray}
where $S=0$ and 1 for singlet and triplet, respectively. The dispersion
coefficients $C_n$ are common to both potentials and are taken from Yan {\em et
al.}\ \cite{Yan1996}. The exchange contribution is \cite{Smirnov:1965}
\begin{equation}
V_{\rm ex}(R) = A (R/a_0)^\gamma \exp(-\beta R/a_0),
\end{equation}
with parameters $\beta=1.25902\ a_0^{-1}$ and $\gamma=4.55988$ \cite{Cote1994}.
The prefactor $A$ is chosen to match the difference between the singlet and
triplet potentials at $R=15.5\ a_0$. The exchange term makes an attractive
contribution for the singlet and a repulsive contribution for the triplet.

The detailed shapes of the short-range singlet and triplet potentials are
relatively unimportant for the ultracold scattering properties and
near-threshold binding energies considered here, although it is crucial to be
able to vary the {\em volume} of the potential wells to allow adjustment of the
singlet and triplet scattering lengths. In the present work we retained the
functional forms used by O'Hara {\em et al.}\ \cite{OHara2002} and Z\"urn {\em
et al.}\ \cite{Zurn2013}, which are based on the short-range singlet potential
of Cot\'e {\em et al.}\ \cite{Cote1994} and the short-range triplet potential
of Linton {\em et al.}\ \cite{Linton1999}. We used the methodology of
Ref.~\cite{Cote1994} to connect the short-range and long-range potentials.

The flexibility needed to adjust the singlet and triplet scattering lengths is
provided by simply adding a quadratic shift to each of the singlet and triplet
potentials inside its minimum,
\begin{equation}
V_S^{\rm shift}(R) = S_S (R-R_{eS})^2 \hbox{\qquad for\qquad} R<R_{eS},
\end{equation}
with $R_{e0}=2.673247$ \AA\ and $R_{e1}=4.173$ \AA.

\section{Fitting interaction potentials \label{sec:fit}}

We have carried out simultaneous fits to experimental results for the
scattering properties and near-threshold bound states of both $^6$Li and
$^7$Li. For $^6$Li the experimental data set was exactly the same as for the
fitting described in ref.\ \cite{Zurn2013}, and was made up of highly precise
bound-state energies $E_{\rm b}$, expressed as frequencies $\nu_{\rm b}=|E_{\rm
b}|/h$, for the (1,2) channel at fields between 720 and 812~G \cite{Zurn2013},
together with transition frequencies between states in the (1,2) and (1,3)
channels at fields between 660 and 690~G \cite{Bartenstein2005} and a precise
measurement of the position of the zero-crossing in the scattering length for
the (1,2) channel \cite{Du2008}. For $^7$Li we fitted to a subset of the
bound-state energies for the (1,1) channel measured by Dyke {\em et al.}\
\cite{Dyke2013} at fields between 725 and 737~G, together with measurements of
a zero-crossing in the (1,1) channel \cite{Pollack2009} and two poles in the
(2,2) channel \cite{Gross2011}. We carried out direct least-squares fitting to
the results of coupled-channels calculations, using the I-NoLLS package
\cite{I-NoLLS} (Interactive Non-Linear Least-Squares), which gives the user
interactive control over step lengths and assignments as the fit proceeds.

As described above, it is not adequate to use the same interaction potentials
for $^6$Li and $^7$Li and to rely on mass-scaling (and changes in hyperfine
parameters) to reproduce the results for both isotopes. We have therefore
chosen to introduce different short-range shift parameters $S_0$ and $S_1$ for
the two isotopes, with the difference between them as explicit fitting
parameters. We define $S_S^{(7)}=S_S^{(6)}+\Delta S_S$ and fit to the 4
parameters $S_0^{(6)}$, $\Delta S_0$, $S_1^{(6)}$ and $\Delta S_1$. In
principle we could also fit to additional parameters such as $C_6$, $C_8$,
etc., as was done for Cs \cite{Berninger:Cs2:2013}, but this was not found to
be necessary to reproduce the threshold results for Li.

\begin{table}[tb]
 \setlength{\tabcolsep}{0mm}
\begin{tabular}{r  l | r  l | r  l  l} \hline\hline
\multicolumn{2}{c|}{$^6$Li}& \multicolumn{2}{c|}{Present fit} & \multicolumn{2}{c}{Experiment} \\ \hline
\multicolumn{2}{c|}{$\nu_{\rm b,12}-\nu_{\rm b,13}+\nu_{\rm ff}$} &   \multicolumn{2}{c|}{83 665.9(8)}
     &  \multicolumn{2}{c}{\,83 664.5(10) } &\ \cite{Bartenstein2005} \\
\multicolumn{2}{c|}{at 661.436 G }& \multicolumn{2}{c|}{} &  \multicolumn{2}{r}{}   \\
\multicolumn{2}{c|}{$\nu_{\rm b,12}-\nu_{\rm b,13}+\nu_{\rm ff}$}  & \multicolumn{2}{c|}{83 297.3(5)} & \multicolumn{2}{c}{\,83 296.6(10)}  &\ \cite{Bartenstein2005}\\
\multicolumn{2}{c|}{at 676.090 G} & \multicolumn{2}{c|}{} & \multicolumn{2}{r}{}  \\
\multicolumn{2}{c|}{$\nu_{\rm b,12}$ at 720.965 G }   &\,127&.115(58)\,			&\;\,127&.115(31)  &\ \cite{Zurn2013}     \\
\multicolumn{2}{c|}{$\nu_{\rm b,12}$ at 781.057 G }   &   14&.103(37)  			&   14&.157(24)    &\ \cite{Zurn2013}     \\
\multicolumn{2}{c|}{$\nu_{\rm b,12}$ at 801.115 G }   &    4&.342(24)  			&    4&.341(50)    &\ \cite{Zurn2013}     \\
\multicolumn{2}{c|}{$\nu_{\rm b,12}$ at 811.139 G }   &    1&.828(16)  			&    1&.803(25)    &\ \cite{Zurn2013}     \\
\multicolumn{2}{c|}{Zero in $a_{12}$ }                &   527&.32(8)  			&  527&.5(2)  	   &\ \cite{Du2008}	\\
\multicolumn{2}{c|}{Narrow pole in $a_{12}$\, }       &   543&.41(12)  			&  543&.286(3)	   &\ \cite{Hazlett2012}	\\
\multicolumn{2}{c|}{$a_{\rm s}$  }                    & 45&.154(2)       &\multicolumn{2}{r}{}   \\
\multicolumn{2}{c|}{$a_{\rm t}$  }                    & \;$-$2113&(2) &\multicolumn{2}{c}{} \\
\hline\hline
\end{tabular}
\begin{tabular}{r  l | r  l | r  l  l} \hline\hline
\multicolumn{2}{c|}{$^7$Li}&  \multicolumn{2}{c|}{Present fit} & \multicolumn{2}{c}{Experiment} \\ \hline
\multicolumn{2}{c|}{Pole in $a_{11}$}  &\,737&.69(2)\,& \multicolumn{2}{c}{} \\
\multicolumn{2}{c|}{Zero in $a_{11}$}  & 543&.64(19) & \multicolumn{2}{c}{\,543.6(1)} & \cite{Pollack2009} \\
\multicolumn{2}{c|}{Pole in $a_{22}$}  & 845&.31(4) & \multicolumn{2}{c}{\,844.9(8)}  & \cite{Gross2011} \\
\multicolumn{2}{c|}{Pole in $a_{22}$}  & 893&.78(4) & \multicolumn{2}{c}{\,893.7(4)}  & \cite{Gross2011} \\
\multicolumn{2}{c|}{$\nu_{\rm b,11}$ at 736.8 G }   &    34&.3(1.0)\,			&40&(3)  & \cite{Dyke2013} \\
\multicolumn{2}{c|}{$\nu_{\rm b,11}$ at 736.5 G }   &    61&.4(1.3)  			&62&(2)  & \cite{Dyke2013} \\
\multicolumn{2}{c|}{$\nu_{\rm b,11}$ at 735.5 G }   &   209&.2(2.3)  			&212&(2) & \cite{Dyke2013} \\
\multicolumn{2}{c|}{$\nu_{\rm b,11}$ at 734.3 G }   &   474&.6(3.5)  			&469&(3) & \cite{Dyke2013} \\
\multicolumn{2}{c|}{$\nu_{\rm b,11}$ at 733.5 G }   &   772&(5)  				&775&(9) & \cite{Dyke2013} \\
\multicolumn{2}{c|}{$\nu_{\rm b,11}$ at 732.1 G }   &  1378&(7)  				&1375&(10) & \cite{Dyke2013} \\
\multicolumn{2}{c|}{$\nu_{\rm b,11}$ at 728.0 G }   &\,4114&(14)  				&4019&(90) & \cite{Dyke2013} \\
\multicolumn{2}{c|}{$a_{\rm s}$  }                  &    34&.331(2) &\multicolumn{2}{c}{} &  \\
\multicolumn{2}{c|}{$a_{\rm t}$  }                  & $-$26&.92(7)  &\multicolumn{2}{c}{} &  \\
\hline\hline
\end{tabular}
\caption{Quality of fit between coupled-channels calculations on the best-fit
4-parameter Li potential and the experiments, together with key derived
quantities calculated using the potential. All frequencies are given in kHz,
all lengths in $a_0$ and all magnetic fields in G. The quantities in
parentheses are statistical 95\% confidence limits in the final digits for the
fit results and quoted uncertainties for the experiments. The quantity
$\nu_{\rm ff}$ is the frequency of the transition between thresholds 2 and 3
for free $^6$Li atoms at the magnetic field concerned.} \label{quality-of-fit}
\end{table}

The set of experimental results used for fitting is listed in Table
\ref{quality-of-fit}. The quantity optimized in the least-squares fits was the
sum of squares of residuals ((obs$-$calc)/uncertainty), with the uncertainties
listed in Table \ref{quality-of-fit}. We carried out both 3-parameter and
4-parameter fits, either including or excluding the $\Delta S_1$ parameter that
describes the deviation from mass-scaling for the weakly bound triplet
potential. We found that 3-parameter fits were capable of reproducing most of
the experimental results, but gave a zero-crossing about 1~G in error for the
scattering length in the (1,1) channel for $^7$Li. The 4-parameter fit, by
contrast, was able to reproduce this along with all the other data. We thus
consider the 4-parameter fit preferable, and give the results based on it in
Table \ref{quality-of-fit}. The optimized parameter values are given in Table
\ref{fitparms}, together with their 95\% confidence limits \cite{LeRoy:1998}.
The optimized potential for $^6$Li is identical to that of Ref.\
\cite{Zurn2013}, because the data set is identical for this isotope, but the
parameter correlations and hence the 95\% confidence limits are different in
the 4-parameter space used in the present work. It may be seen that the 95\%
confidence limits for the $\Delta S_0$ parameter is less than 2.5\% of its
value, and even that for the $\Delta S_1$ parameters is less than 25\% of its
value.

\begin{table}[tb]
\caption{Parameters of the fitted potential, together with statistical
confidence limits and sensitivities that indicate the precision needed to
reproduce the calculated quantities.} \label{fitparms}
\begin{tabular}{l|rrr}
\hline\hline
& fitted value & confidence   & sensitivity \\
&              & limit (95\%) &             \\
\hline
$S_0^{(6)}$  ($\mu E_{\rm h}\,a_0^{-2}$) & $-11.8959$  & $0.0383$   & $0.0003$ \\
$\Delta S_0$ ($\mu E_{\rm h}\,a_0^{-2}$) &   $3.0429$  & $0.0714$   & $0.0006$ \\
$S_1^{(6)}$  ($\mu E_{\rm h}\,a_0^{-2}$) &   $0.51031$ & $0.00203$  & $0.00002$ \\
$\Delta S_1$ ($\mu E_{\rm h}\,a_0^{-2}$) &   $0.69955$ & $0.15149$  & $0.00142$ \\
\hline\hline
\end{tabular}
\end{table}

It should be emphasized that the 95\% confidence limits are {\em statistical}
uncertainties within the particular parameter set. They do not include any
errors due to the choice of the potential functions. Such model errors are far
harder to estimate, except by performing a large number of fits with different
potential models, which is not possible in the present case.

In a correlated fit, the statistical uncertainty in a fitted parameter depends
on the degree of correlation. However, to reproduce the results from a set of
parameters, it is often necessary to specify many more digits than implied by
the uncertainty. A guide to the number of digits required is given by the {\em
parameter sensitivity} \cite{LeRoy:1998}, which essentially measures how fast
the observables change when one parameter is varied with all others held fixed.
This quantity is included in Table \ref{fitparms}.

The singlet and triplet scattering lengths and the pole positions of the
$s$-wave resonances are not directly observed quantities. Nevertheless, their
values may be extracted from the final potential. In addition, the statistical
uncertainties in any quantity obtained from the fitted potentials may be
obtained as described in Ref.\ \onlinecite{LeRoy:1998}. The values and 95\%
confidence limits obtained in this way, for both the derived parameters such as
scattering lengths and the experimental observables themselves, are given in
Table \ref{quality-of-fit}. It may be noted that the statistical uncertainties
in the calculated properties are independent of the experimental uncertainties,
and in some cases are smaller.

\section{Discussion of Results}

\subsection{Born-Oppenheimer corrections}

For a single potential curve with long-range form $-C_6R^{-6}$, the scattering
length is given semiclassically by \cite{Gribakin1993}
\begin{equation}
a=\bar{a} \left[ 1- \tan \left( \Phi- \frac{\pi}{8} \right) \right],
\label{eq:gf}
\end{equation}
where $\Phi$ is a phase integral evaluated at the threshold energy,
\begin{equation}
\Phi=\int_{R_0}^\infty \left(\frac{-2\mu V(R)}{\hbar^2}\right)^\frac{1}{2}\,dR
\end{equation}
and $R_0$ is the inner turning point at this energy. A value of $a$ thus
directly implies the fractional part of $\Phi/\pi$. In the Born-Oppenheimer
approximation, $V(R)$ is independent of reduced mass, so that the values of
$\Phi/\pi$ for different isotopologues are related by simple mass scaling. For
alkali metals heavier than Li, such mass scaling is very accurate
\cite{Seto:2000, Docenko:2007, Falke:2008, Simoni:2008, Cho:RbCs:2013,
Blackley:85Rb:2013}. For Li, however, significant corrections are needed, as
shown in Section \ref{sec:fit}. Eq.\ (\ref{eq:gf}) may be used to convert the
singlet and triplet scattering lengths in Table \ref{quality-of-fit} into the
corresponding fractional parts of $\Phi/\pi$. Together with the reduced mass
ratio $\mu^{(7)}/\mu^{(6)}=1.1166394$, these are sufficient to determine
unambiguously the integer part of $\Phi/\pi$, which for $^6$Li is 38 for the
singlet state and 10 for the triplet state. The resulting values for $\Phi/\pi$
for both isotopes are given in Table \ref{tab:phi}. Comparison of these values
obtained directly from the scattering lengths for $^7$Li with those obtained by
mass-scaling the $^6$Li results shows that the non-Born-Oppenheimer terms
contribute an additional $-9.4\times10^{-4}$ to $\Phi/\pi$ for singlet $^7$Li
and $-1.7\times10^{-3}$ for triplet $^7$Li.  The deviations from mass-scaling
in the scattering lengths thus do not by themselves contain any information on
the $R$-dependence of the non-Born-Oppenheimer terms, but do provide a strong
constraint on their overall magnitude, which could be included in spectroscopic
fits such as those of refs.\ \cite{Leroy2009} and \cite{Dattani2011}.

\begin{table}[tb]
\caption{Phase integrals obtained from the fitted singlet and triplet
scattering lengths.} \label{tab:phi}
\begin{tabular}{l|lll}
\hline\hline
& $^6$Li & mass-scaled & \quad $^7$Li \\
&        & $^6$Li to $^7$Li & \\
\hline
$\Phi_{\rm s}/\pi$ & 38.97463 \quad & 42.09250 & 42.09156 \\
$\Phi_{\rm t}/\pi$ & 10.62056 & 11.47018 & 11.46846 \\
\hline\hline
\end{tabular}
\end{table}

\subsection{Relationship of binding energy to scattering length}

\begin{figure}[tb]
\includegraphics[width=\columnwidth,clip]{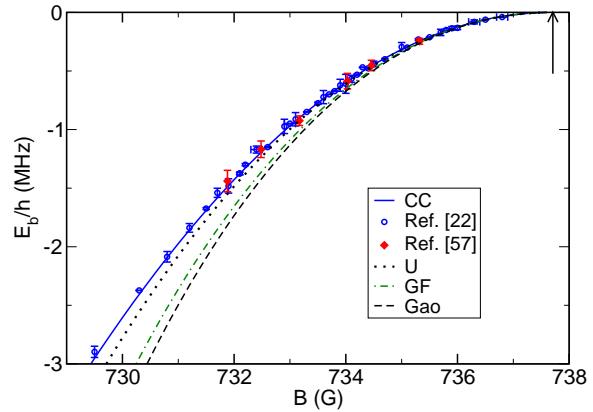}
\caption{(Color online) Observed and calculated bound-state energies $E_\mathrm{b}$
in the (1,1) channel of $^7$Li as a function of magnetic field $B$:
Coupled-channels calculation (solid line), experimental results from
Refs.~\cite{Dyke2013} (open circles) and~\cite{Navon2011} (diamonds), and
approximations from the universal (dotted), Gribakin-Flambaum (dot-dash) and
Gao (dashed) formulas. } \label{fig:NormalCC+Data}
\end{figure}

Figure~\ref{fig:NormalCC+Data} shows the calculated binding energy as a
function of $B$ for the (1,1) channel of $^7$Li, illustrating the good
agreement with the results of Dyke {\it et al.}~\cite{Dyke2013}.  Our
calculation also agrees well with the results of Navon {\it et
al.}~\cite{Navon2011}, although we did not use these in the fits to obtain
potentials.  We do not show the results of Gross {\it et al.}~\cite{Gross2011},
since they report a magnetic-field calibration uncertainty of 0.3~G.  Their
results would be in reasonable agreement with our calculation if they were
shifted to lower field by about 0.34~G.

Figure~\ref{fig:NormalCC+Data} also shows the results of three simple
single-channel formulas that relate the binding energy to the scattering
length,
\begin{eqnarray}
 E^\mathrm{U}_\mathrm{b} &=& -\frac{\hbar^2}{2\mu a^2}  \label{eq:U}, \\
 E^\mathrm{GF}_\mathrm{b} &=& -\frac{\hbar^2}{2\mu (a-\bar{a})^2},  \label{eq:GF} \\
E^\mathrm{Gao}_\mathrm{b} &=& -\frac{\hbar^2}{2\mu (a-\bar{a})^2}
\left [  1 + \frac{g_1 \bar{a}}{a-\bar{a}} + \frac{g_2 \bar{a}^2}{(a - \bar{a})^2} \right ],  \label{eq:G}
 \end{eqnarray}
where $g_1 = \Gamma(1/4)^4/(6\pi^2) \approx 2.9179$ and $g_2=(5/4)g_1^2-2
\approx -0.9468$. The first formula is the familiar universal relationship
between the last bound state and the scattering length~\cite{Chin2010}.  The
second gives a correction for the van der Waals potential that follows from the
work of Gribakin and Flambaum~\cite{Gribakin1993}.  The final formula, due to
Gao~\cite{Gao2004b}, includes higher-order corrections based on the analytic
solutions of the Schr{\"o}dinger equation for the van der Waals potential.

A good way to highlight the differences between the calculations, experimental
results and approximate formulas is to plot $\epsilon r^2$ as a function of
$1/r^2$, using van der Waals units of energy $\epsilon = E_\mathrm{b}/\bar{E}$
and length $r=a/\bar{a}$ \footnote{In linear plots, it is advantageous to show
$\epsilon r^2$ as a function of $1/r$ or $1/r^2$ to compress the region near a
pole in $r$. Here, however, we show $1/r^2$ on a logarithmic axis, and in this
case it is equivalent to use either $r$ or $1/r$. For ease of conversion, we
show the $r$ scale on the upper axes.}.  In these units the approximate
formulas are
\begin{eqnarray}
\epsilon^\mathrm{U}r^2   &=& -1  \label{eq:Uvdw}, \\
\epsilon^\mathrm{GF}r^2  &=& \frac{-1}{\left(1-\frac{1}{r}\right)^2},  \label{eq:GFvdw} \\
\epsilon^\mathrm{Gao}r^2 &=& \frac{-1}{\left(1-\frac{1}{r}\right)^2}
\left [  1 + \frac{g_1}{r-1} + \frac{g_2 }{(r - 1)^2} \right ].  \label{eq:Gvdw}
\end{eqnarray}
In this representation, the positions of the experimental points themselves
depend upon the $a(B)$ mapping used to interpret them.

\begin{figure}[tb]
\includegraphics[width=\columnwidth,clip]{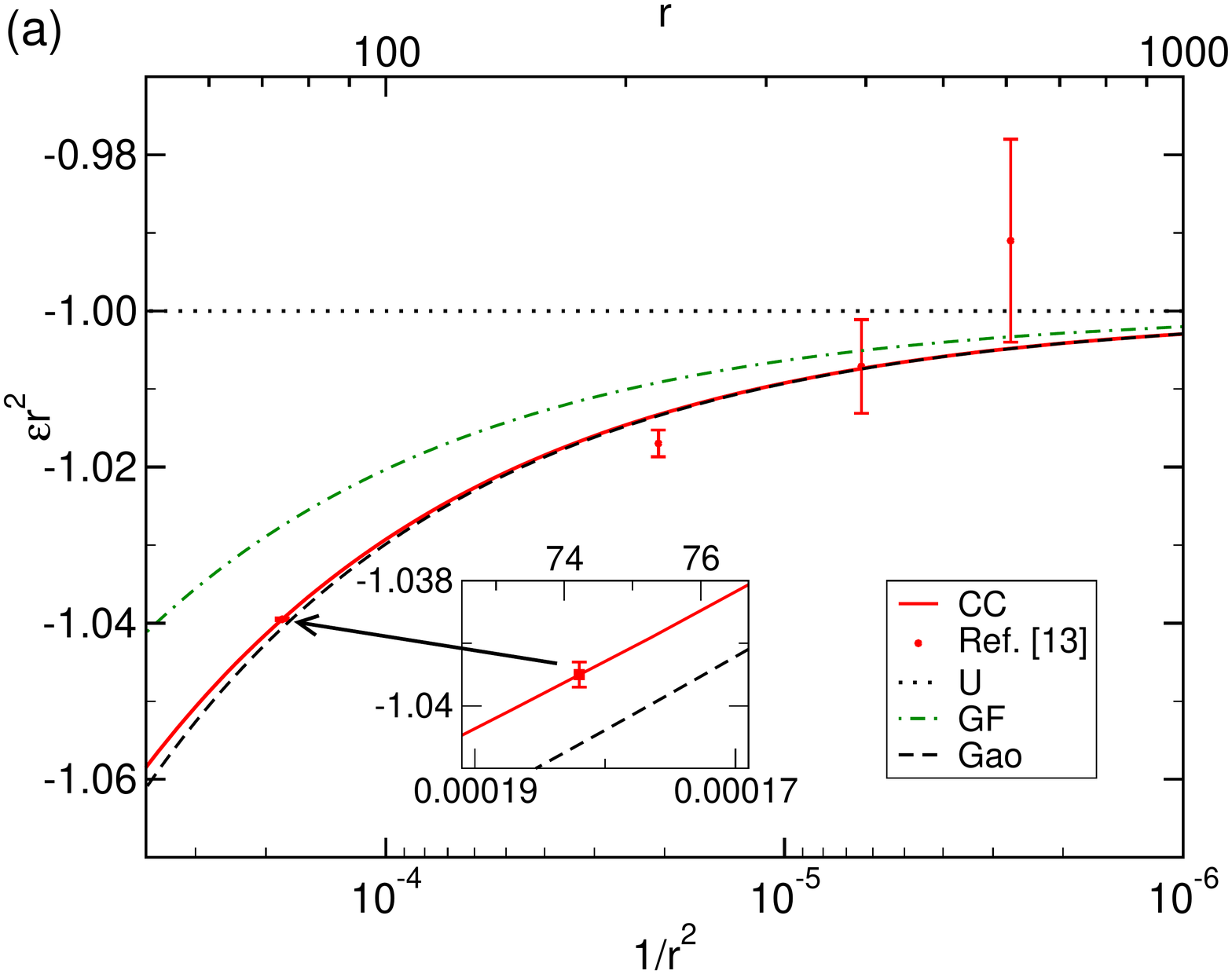}
\includegraphics[width=\columnwidth,clip]{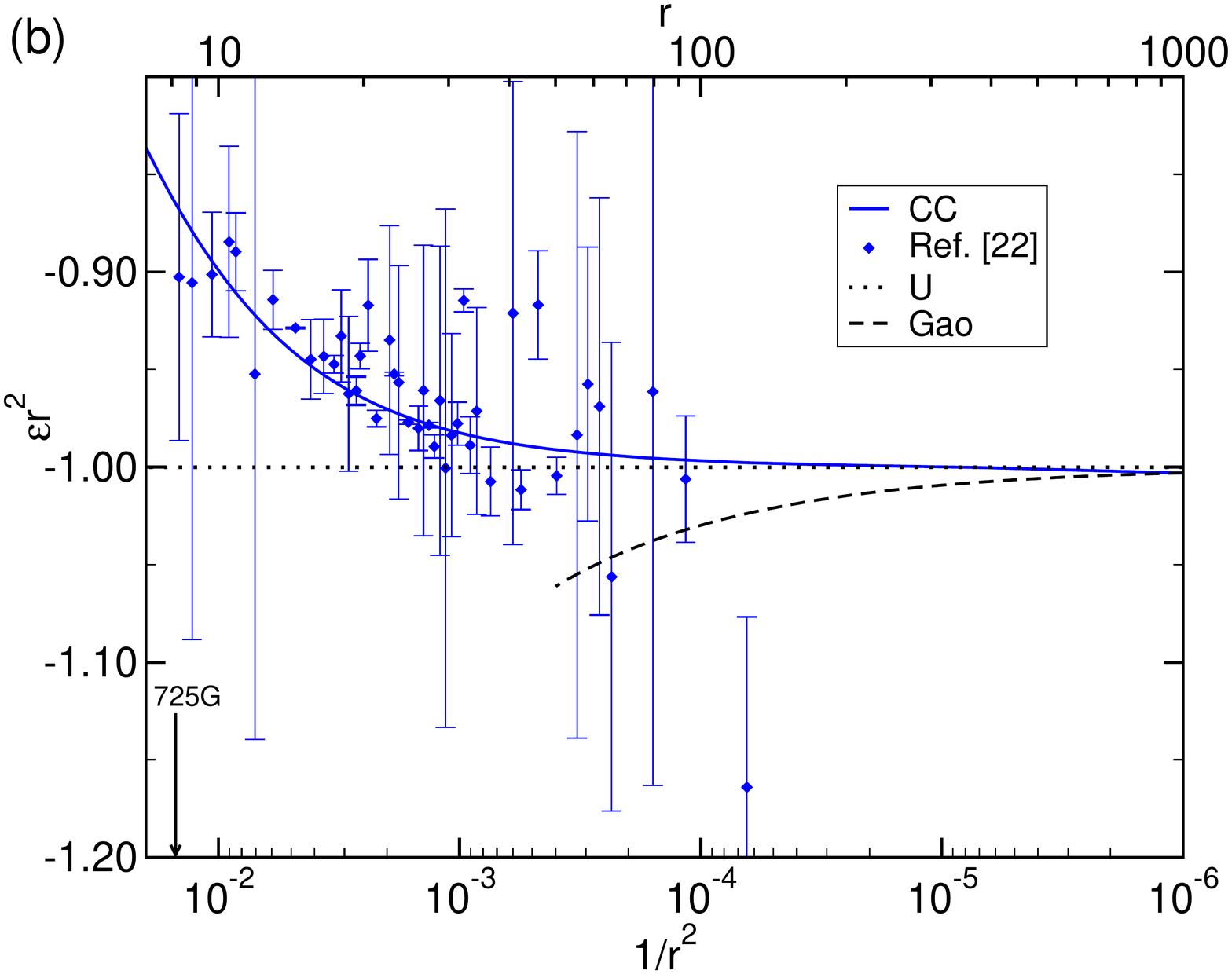}
\caption{(Color online) Plots of $\epsilon r^2$ against $1/r^2$, using van der
Waals units of energy and length.  The solid lines show the coupled-channels
calculation, the points give the experimental results mapped using the
calculated $a(B)$ function evaluated at the measured $B$ values, and the
dotted, dot-dash, and dashed lines give the results of the universal,
Gribakin-Flambaum, and Gao formulas, respectively.  (a) Bound-state energies in
the (1,2) channel of $^6$Li, showing experimental results from Z\"urn {\it et
al.}~\cite{Zurn2013}.  The inset shows an expanded view near $1/r^2=0.00018$,
or $a \approx 2200$ $a_0$. (b) Bound-state energies in the (1,1) channel of
$^7$Li, showing experimental results from Dyke {\it et al.}~\cite{Dyke2013}. }
\label{fig:vdwCompare}
\end{figure}

\begin{figure}[tb]
\includegraphics[width=\columnwidth,clip]{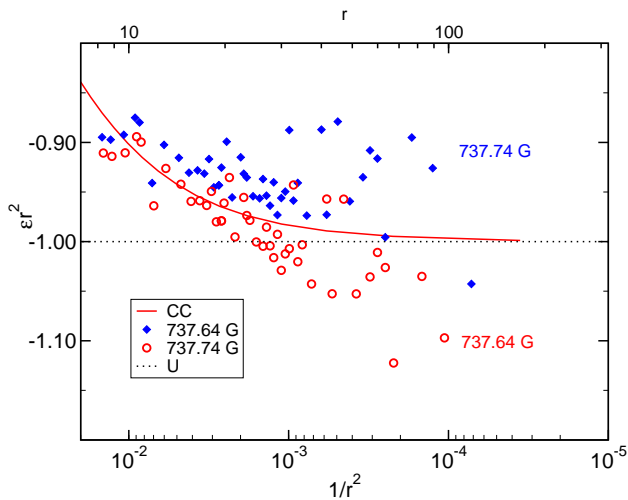}
\caption{(Color online) Plots of $\epsilon r^2$ against $1/r^2$ in the (1,1)
channel of $^7$Li, showing the experimental results of Dyke {\it et
al.}~\cite{Dyke2013}, mapped with two different $a(B)$ functions calculated
when the $^1\Sigma_g^+$ potential is adjusted slightly to give two different
pole positions, 737.74~G and 737.64~G, which differ by $\pm 0.05$~G from our
fitted value of 737.69(2)~G.   } \label{fig:offpole}
\end{figure}

Figure~\ref{fig:vdwCompare}(a) compares the coupled-channels bound-state
energies, the experimental results of Z\"urn {\it et al.}~\cite{Zurn2013} and
the three approximate formulas near the pole of the broad resonance in the
(1,2) channel of $^6$Li. Even for this open-channel-dominated resonance with
$s_\mathrm{res} \gg 1$, the universal formula differs from the coupled-channel
bound-state energy by 3\% at $1/r^2=0.0001$, or $a = 100 \bar{a}$. The
Gribakin-Flambaum correction reduces the difference to about 1\%, while Gao's
single-channel formula, Eq.~(\ref{eq:Gvdw}), gives an excellent representation
of the bound-state energy that differs from the coupled-channels result by only
0.06\% at $1/r^2=0.0001$. Nevertheless, the measurement precision of ref.\
\cite{Zurn2013} is so good that the difference between the experimental results
and the Gao formula reaches 5 times the experimental error of 0.02\% near
$1/r^2=0.00018$, or $a = 74 \bar{a}$, as shown in the inset of Fig.\
\ref{fig:vdwCompare}(a). The coupled-channels model, on the other hand, agrees
with the measured value within experimental uncertainty. It may be noted that
the universal value of $-1$ for $\epsilon r^2$ differs from the experimental
value by nearly 200 times the experimental uncertainty for this point.

Figure~\ref{fig:vdwCompare}(b) shows a similar comparison between the
calculated bound-state energies and the experimental results of Dyke {\it et
al.}~\cite{Dyke2013} for the (1,1) channel of $^7$Li. There is much more
scatter in the experimental results than for the $^6$Li results of Z\"urn {\it
et al.}, but the overall agreement with the coupled-channels model is good. The
fluctuations in the experimental results near the pole are much more evident in
this plot than in Fig.\ \ref{fig:NormalCC+Data}. It is also clear that the
bound-state energies are quite poorly represented by Gao's single-channel
formula around this closed-channel-dominated resonance with $s_\mathrm{res} <
1$. The apparent agreement with the universal formula over a wider range than
for $^6$Li is an artifact, arising because, in the $^7$Li case, $\epsilon r^2$
deviates from its universal value of $-1$ in the opposite direction to that
predicted by Gao's single-channel formula.

Plotting measured values of $E_{\rm b}(B)$ as $\epsilon r^2$ against $1/r^2$
provides a sensitive test of the mapping $a(B)$ between magnetic field $B$ and
scattering length $a$. In particular, if a mapping with an incorrect pole
position is used, the results do not properly approach the limit $-1$ as $1/r^2
\to 0$. Figure \ref{fig:offpole} shows this behavior for two coupled-channels
models that have slightly incorrect pole positions, in this case differing by
about $\pm 0.05$~G from our best value. It may be seen that the two sets of
points clearly deviate from $-1$ in opposite directions as $1/r^2 \to 0$. Such
an error would be even more apparent in Fig.~\ref{fig:vdwCompare}(a) if we
plotted the recent experimental results of Z\"urn {\it et al.}~\cite{Zurn2013}
using the older $a(B)$ mapping from Bartenstein {\it et
al.}~\cite{Bartenstein2005}. The points would then lie well off the
coupled-channels and Gao curves, and the point nearest the pole would lie 10\%
away from the limiting value of $-1$ because of the 2~G difference in pole
position between Bartenstein {\it et al.} and Z\"urn {\it et al.}

\section{Conclusions}

We have produced a new coupled-channels model for the collision of two cold Li
atoms, based on published bound-state and scattering properties for both
isotopic species $^6$Li and $^7$Li.  Our new model simultaneously fits results
for both isotopes to obtain $^1\Sigma_g^+$ and $^3\Sigma_u^+$ interaction
potentials that include terms to account for the small mass-dependent
corrections to the Born-Oppenheimer approximation.  It is necessary to include
such corrections for both the $^1\Sigma_g^+$ and $^3\Sigma_u^+$ potentials.
Our calculations show the overall magnitude of the mass-correction effect
through the difference in threshold phase for each of these potentials. Such
constraints on the magnitude can be tested against theory if future {\it ab
initio} studies determine the magnitude of the mass-dependent corrections to
the Born-Oppenheimer approximation.

Our new model allows a careful study of the differences between three different
approximate formulas that relate the binding energy of the Feshbach molecule to
the scattering length of the two atoms.  To do this, it is helpful to express
the binding energy and scattering length in van der Waals units of $\bar{E}$
and $\bar{a}$.  In these units, the conventional universal relationship
predicts that the product of the bound-state energy and the square of
scattering length has a constant magnitude $-1$, independent of species.
Departures from this limit are readily apparent. Plotting the results in this
way requires a mapping between the magnetic field $B$ and the scattering length
$a$ and provides a sensitive test of the position of the resonance pole in
$a(B)$. The high precision of measurement for $^6$Li$_2$ bound states and the
quality of our coupled-channels model permit the differences between the
experimental results and the approximate formulas to be seen clearly.  Both the
$^6$Li$_2$ and $^7$Li$_2$ bound states show pronounced departure from the
universal formula as the binding energy increases away from resonance, even
while the scattering length remains large compared to the characteristic van
der Waals length.   The two isotopologues show quite different patterns of
variation away from the universal relationship, due to the large difference in
their resonance pole strength $s_\mathrm{res}$.  The binding energies near the
strong $^6$Li resonance with $s_\mathrm{res} \gg 1$ agree well with the Gao
relationship~\cite{Gao2004b} for a single channel. However, those near the much
weaker $^7$Li resonance with $s_\mathrm{res} < 1$ are not well reproduced by
any of the approximate formulas and require a coupled-channels model.

\begin{acknowledgments}
The authors thank Professors Randall Hulet and Christophe Salomon for providing
numerical tables of their published experimental results and acknowledge
support from AFOSR-MURI FA9550-09-1-0617 and EOARD Grant FA8655-10-1-3033.
\end{acknowledgments}

\bibliography{../Allrefs_psj,../all,psj}

\end{document}